\documentclass[11pt]{article}
\usepackage{graphics}
\usepackage{epsfig}
\usepackage{amsmath}
\usepackage{amssymb}
\usepackage{citesort}

\makeatletter

\def\dfrac#1#2{{\displaystyle {#1 \over #2}}}
\def\dsum{\mathop{\displaystyle \sum }}

\newcommand{\be}{\begin{equation}}
\newcommand{\ee}{\end{equation}}
\newcommand{\bea}{\begin{eqnarray}}
\newcommand{\eea}{\end{eqnarray}}
\newcommand{\nn}{\nonumber}
\newcommand{\msb}{\overline{\rm{MS}}}
\newcommand{\mev}{\,{\rm MeV}}   
\newcommand{\gev}{\,{\rm GeV}}   
\newcommand{\ri}{\mbox{RI-MOM}}
\makeatother
\begin{document}
\thispagestyle{empty}
\begin{flushright}
\scalebox{.9}{\small 
\begin{tabular}{l}
{\tt FTUV-05-0926}\\
{\tt IFIC/05-43}\\
{\tt LPT Orsay 05-61}\\
{\tt RM3-TH/05-6}\\
\end{tabular} 
}\end{flushright}
\vskip 2.0cm\par

\begin{center}
{\par\centering \textbf{\Large Non-perturbatively renormalised light quark
masses}}\\ \vskip 0.3cm
{\par\centering \textbf{\Large from a lattice simulation with \boldmath{$N_f=2$}}}\\
\vskip 0.75cm\par
{\par\centering \sc D.~Be\'cirevi\'c$^a$, B.~Blossier$^a$, Ph.~Boucaud$^a$,
V.~Gim\'enez$^b$, \\ V.~Lubicz$^{c,d}$, F.~Mescia$^{c,e}$, S.~Simula$^d$ 
and C.~Tarantino$^{c,d}$}
\vskip 0.3cm\par
{\par\centering \large SPQ$_\mathrm{CD}$R Collaboration}
{\vskip 0.25cm\par}
\end{center}
{\par\centering 
\textsl{$^a$Laboratoire de Physique Th\'eorique (B\^at.210), Universit\'e
Paris Sud,}\\
\textsl{Centre d'Orsay, F-91405 Orsay-cedex, France.}\\
\vskip 0.3cm\par}
{\par\centering \textsl{$^b$Dep.de F\' \i sica Te\`orica and IFIC, Univ.~de 
Val\`encia,}\\
\textsl{Dr.~Moliner 50, E-46100 Burjassot, Val\`encia, Spain.}\\
\vskip 0.3cm\par}
{\par\centering \textsl{$^c$Dipartimento di Fisica, Universit\`a di Roma
Tre, }\\
\textsl{Via della Vasca Navale 84, I-00146 Rome, Italy.}\\
\vskip 0.3cm\par}
{\par\centering \textsl{$^d$INFN, Sezione di Roma III, Via della Vasca
Navale 84, I-00146 Rome, Italy.}
\vskip 0.3cm\par}
{\par\centering \textsl{$^e$INFN, Lab. Nazionali di Frascati, Via E. Fermi
40, I-00044 Frascati, Italy.}
\vskip 0.3cm\par}
\begin{abstract}
We present results for the light quark masses obtained from a lattice QCD 
simulation with $N_f=2$ degenerate Wilson dynamical quark 
flavours. 
The sea quark masses of our lattice, of spacing $a\simeq 0.06$~fm,  
are relatively heavy, i.e., they cover the range corresponding to 
$0.60 \lesssim M_P/M_V \lesssim 0.75$. 
After implementing the non-perturbative $\ri$ method to renormalise quark 
masses, we obtain $m_{ud}^{\msb}(2\gev)=4.3\pm 0.4^{+1.1}_{-0}~\mev$, and 
$m_s^{\msb}(2\gev)=101\pm 8^{+25}_{-0}~\mev$, which are about $15$\%
larger than they would be if renormalised perturbatively. 
In addition, we show that the above results are compatible with 
those obtained in a quenched simulation with a similar lattice.
\end{abstract}

\renewcommand{\thefootnote}{\arabic{footnote}}
\newpage
\setcounter{page}{1}
\setcounter{footnote}{0}
\setcounter{equation}{0}
\setcounter{table}{0}

\section{Introduction}
An accurate determination of quark masses is highly important for both 
theoretical studies of flavour physics and particle physics phenomenology. 
Within the quenched approximation ($N_f=0$), lattice QCD calculations 
reached a few percent accuracy, so that the error due to the use of quenched 
approximation became the main source of uncertainty. 
To examine the effects of the inclusion of the 
sea quarks several lattice results with either $N_f=2$ or $N_f=3$ dynamical 
quarks have  recently been reported
~\cite{Rakow,AliKhan,Aoki,Namekawa,Dawson,Ishikawa,Aubin,Gockeler,DellaMorte}. 
Many of them seem to indicate that the unquenching lowers the physical values of
light quark masses. 
For example, the strange quark mass, which  in the quenched 
approximation is about $m_s^{\msb}(2\gev)\simeq 100$~MeV, becomes  
$m_s^{\msb}(2\gev)\simeq 85\div 90$~MeV when $N_f=2$ dynamical quarks are 
included~\cite{AliKhan,Aoki}, and even smaller with $N_f=3$, namely 
$m_s^{\msb}(2\gev)\simeq 75\div 80$ MeV~\cite{Ishikawa,Aubin}.

The accuracy of the results obtained from simulations with $N_f\neq 0$, however,
is still limited by several systematic uncertainties. 
In particular, precision quenched studies of light quark masses evidenced 
the importance of computing the mass renormalisation constants 
non-perturbatively. 
In almost all the studies with dynamical fermions performed so far, however, 
renormalisation has been made by means of one-loop boosted 
perturbation theory (BPT). 
Only very recently, non-perturbative renormalisation (NPR) has been implemented
in a simulation with $N_f=2$ by the QCDSF collaboration. 
The resulting strange quark mass value is $m_s^{\msb}(2\gev)= 119\pm9$~MeV~
\cite{Gockeler}, thus larger by about 
$20$\% than the one previously reported in ref.~\cite{Aoki}, where 
the same lattice action and the same vector definition of the bare quark mass 
have been used, but with the renormalisation constant estimated perturbatively. 
Recently, also the Alpha collaboration reported their value for the NPR-ed 
strange quark mass, $m_s^{\msb}(2\gev)= 97\pm 22$~MeV~\cite{DellaMorte}. 
It appears that the results of both QCDSF and Alpha are consistent with their 
previous quenched estimates~\cite{Gockeler2,Garden}.

In this paper we present our results for the light quark masses obtained on a 
$24^3\times 48$ lattice with $N_f=2$ dynamical mass-degenerate quarks, in which 
we implement the $\ri$ NPR method of refs.~\cite{Martinelli,Becirevic}. 
In the numerical simulation we used the Hybrid Monte Carlo 
algorithm (HMC)~\cite{Duane,HMC}, and chose to work with the Wilson
plaquette gauge action and the Wilson quark action 
at $\beta = 5.8$. The corresponding lattice spacing is $a\simeq 0.06$~fm, 
and the spatial extension of our lattice is $L\simeq 1.5$~fm. Gauge configurations have been 
generated with four different values of the sea quark mass, for which the ratio of the pseudoscalar 
over vector meson masses lies in the range $M_P/M_V \simeq 0.60 \div 0.75$.  
Our final results for the average up/down and strange quark masses are
\be
\label{eq:res1}
\renewcommand{\arraystretch}{1.5}
\begin{array}{ll}
m_{u/d}^{\msb}(2\gev)=4.3(4)(^{+1.1}_{-0})~\mev \\
m_s^{\msb}(2\gev)=101(8)(^{+25}_{-0})~\mev\,
\end{array}\qquad
\left(\begin{array}{cc} N_f=2 \\ \ri\end{array}\right) \,,
\renewcommand{\arraystretch}{1.0}
\ee
where the first error is statistical and the second systematic, the latter 
dominated by discretisation lattice artifacts. 
Our central values refer to the quark mass definition based on the use of the 
axial Ward identity. 
The above results are larger than the ones we obtain by using one-loop BPT 
to compute the mass renormalisation, namely 
\be
\label{eq:res2}
\renewcommand{\arraystretch}{1.5}
\begin{array}{ll}
m_{ud}^{\msb}(2\gev)=3.7(3)(^{+1.3}_{-0})\mev \\
m_s^{\msb}(2\gev)=88(7)(^{+30}_{-0})\mev \,
\end{array}\qquad 
\left(\begin{array}{cc} N_f=2 \\ \mbox{BPT}\end{array}\right)\,.
\renewcommand{\arraystretch}{1.0}
\ee
The latter values agree with the $N_f=2$ results reported in 
refs.~\cite{AliKhan,Aoki} in which BPT renormalisation has been used,
while the values in eq.~(\ref{eq:res1}) agree with
refs.~\cite{Gockeler,DellaMorte} in which NPR
has been implemented. 
The values given in eq.~(\ref{eq:res1}) can be also compared with our quenched 
estimate,  
\be
\renewcommand{\arraystretch}{1.5}
\begin{array}{ll}
m_{ud}^{\msb}(2\gev)=4.6(2)(^{+0.5}_{-0})\mev\, \\ 
m_s^{\msb}(2\gev)=106(2)(^{+12}_{-0})\mev \,
\end{array}\qquad 
\left(\begin{array}{cc} N_f=0 \\ \ri\end{array}\right)\,,
\label{eq:res3}
\renewcommand{\arraystretch}{1.0}
\ee
obtained on a lattice similar in size and resolution, and by employing the same 
Wilson quark action.
We therefore conclude that, within our statistical and systematic accuracy, and 
with the sea quark masses used in our simulations,  we do not observe any 
significant effect due to the presence of dynamical quarks.  
We plan to pursue this study by exploring lighter sea quarks.

The remainder of this paper is organised as follows: in sec.~\ref{sec:simulation} we discuss the details 
of our numerical simulations; in sec.~\ref{sec:masses} we present the
results for the meson masses and the bare quark masses, directly accessed
from the simulation; in sec.~\ref{sec:NPR} we provide the mass
renormalisation factors, while in sec.~\ref{sec:result} we determine the
physical quark mass values; we briefly conclude in 
sec.~\ref{sec:concl}.

\section{\label{sec:simulation}Simulation details}
In our numerical simulation we choose to work with the Wilson action
\be
S_W = S_g+S_q\,,
\ee
where $S_g$ is the standard Wilson plaquette gauge action,
\be
S_g=\dfrac{\beta}{6} \dsum_{p} \mathrm{Tr}\, U_{p}=\dfrac{\beta}{6} 
\dsum_{x,\mu\nu}\mathrm{Tr}\,
[U_{x,\mu}U_{x+\hat\mu, \nu}U_{x+\hat\nu, \mu}^{\dagger}U_{x, \nu}^{\dagger}]\,,
\ee
and $N_f=2$ flavours of mass-degenerate dynamical quarks are included by using 
the Wilson action
\be
S_q = \dsum_{x,y}\bar{q}_x D_{xy} q_y = \dsum_{x,y}\bar{q}_x \left[
\delta_{xy} - \kappa \dsum_{\mu} [(1-\gamma_\mu)U_{x,\mu}\delta_{x+\hat\mu,y}+
(1+\gamma_\mu)U_{x,\mu}^\dagger\delta_{x,y+\hat\mu}] \right] q_y \,,
\ee
with $\kappa$ being the usual hopping parameter.  The simulation parameters used
 in this work are summarised in table~\ref{TTtable1}.
\begin{table}[h!]
\centering
\scalebox{.92}{ 
\begin{tabular}{|c|cccc|ccc|}
\cline{2-8}
\multicolumn{1}{l|}{}&\multicolumn{4}{|c|}{
{\phantom{\Huge{l}}} \raisebox{-.2cm} {\phantom{\Huge{j}}}
$\mathbf \beta = 5.8$}&
\multicolumn{3}{|c|}{$\mathbf \beta = 5.6$} \\ 
\hline
{\phantom{\Large{l}}}\raisebox{-.1cm}{\phantom{\Large{j}}}
$\kappa_s$ & $0.1535$ & $0.1538$ & $0.1540$ & $0.1541$& $0.1560$ & $0.1575$ & $0.1580$\\
\hline
\hline

$L^3\times T$ &\multicolumn{4}{|c|}{
{\phantom{\Large{l}}}\raisebox{-.1cm}{\phantom{\Large{j}}}
$24^3 \times 48$}&\multicolumn{3}{|c|}{
$16^3 \times 48$ }\\ 
\hline
{\phantom{\Large{l}}}\raisebox{-.1cm}{\phantom{\Large{j}}}
minutes/traj. & $10$ & $12$ & $16$ & $21$ &$5.3$ &$13$ &$28$\\
{\phantom{\Large{l}}}\raisebox{-.1cm}{\phantom{\Large{j}}}
$T_{MC}$ & $2250$ & $2250$ & $2250$ & $2250$ & $2250$& $2250$& $2250$ \\
{\phantom{\Large{l}}}\raisebox{-.1cm}{\phantom{\Large{j}}}
$N_{conf}$ & $50$ & $50$ & $50$ & $50$& $50$& $50$& $50$\\
{\phantom{\Large{l}}}\raisebox{-.1cm}{\phantom{\Large{j}}}
$\langle P\rangle$ & $0.59268(3)$ & $0.59312(2)$ & $0.59336(3)$ & $0.59341(3)$ & $0.56990(6)$& $0.57248(6)$& $0.57261(6)$ \\
\hline

$L^3\times T$ &\multicolumn{4}{|c|}{
{\phantom{\Large{l}}}\raisebox{-.1cm}{\phantom{\Large{j}}}
$16^3 \times 48$} \\ 
\cline{1-5}
{\phantom{\Large{l}}}\raisebox{-.1cm}{\phantom{\Large{j}}}
minutes/traj. & $3.3$ & $4.4$ & $4.4$ & $5.6$\\
{\phantom{\Large{l}}}\raisebox{-.1cm}{\phantom{\Large{j}}}
$T_{MC}$ & $4500$ & $4500$ & $4500$ & $4500$\\
{\phantom{\Large{l}}}\raisebox{-.1cm}{\phantom{\Large{j}}}
$N_{conf}$ & $100$  & $100$ & $100$ & $100$\\
{\phantom{\Large{l}}}\raisebox{-.1cm}{\phantom{\Large{j}}}
$\langle P\rangle$ & $0.59283(3)$ & $0.59314(3)$ & $0.59340(3)$ & $0.59345(3)$\\
\cline{1-5}

\end{tabular} }
\caption{\small \sl Run parameters for the simulations with $N_f=2$. 
From the entire ensemble of generated HMC trajectories of unit length,
$T_{MC}$, we select $N_{conf}$ for the data analysis. 
The stopping condition in the quark matrix inversion ($D_{xy}G_y=B_x$) is 
$||DG-B||<10^{-15}$. $\langle P\rangle$ is the value of the average
plaquette. ``Minutes/traj." refers to the time required to produce one 
trajectory on the APEmille machine ($128$~GFlop).\label{TTtable1}}
\end{table}
To generate the ensembles of gauge field configurations
we employ the HMC algorithm~\cite{Duane,HMC} and implement the LL-SSOR 
preconditioning~\cite{Fischer}. 
In solving the molecular dynamics equations we use the leapfrog
integration scheme. 
Each trajectory of unit length ($\tau =1$) is divided into 
$N_{\rm step}=250$ time-steps.
The resulting $\delta t=\tau/N_{\rm step}=4\times 10^{-3}$ appears to be small 
enough to avoid large energy differences 
along a trajectory, thus increasing the acceptance probability. 
We find  an acceptance probability of about $85$\%. 
To check the reversibility along the trajectory we verify that the quantity 
$||U^\dagger U -1||$ remains unchanged at the relative level of $10^{-11}$. 
For the quark matrix inversion algorithm 
we used the BiConjugate Gradient Stabilized (BiCGStab) method~\cite{bicgstab}. 
The inversion rounding errors are of the order of $10^{-8}$. 

\begin{figure}[t]
\centering
\begin{tabular}{cc}
\epsfxsize=7.5cm
\epsfbox{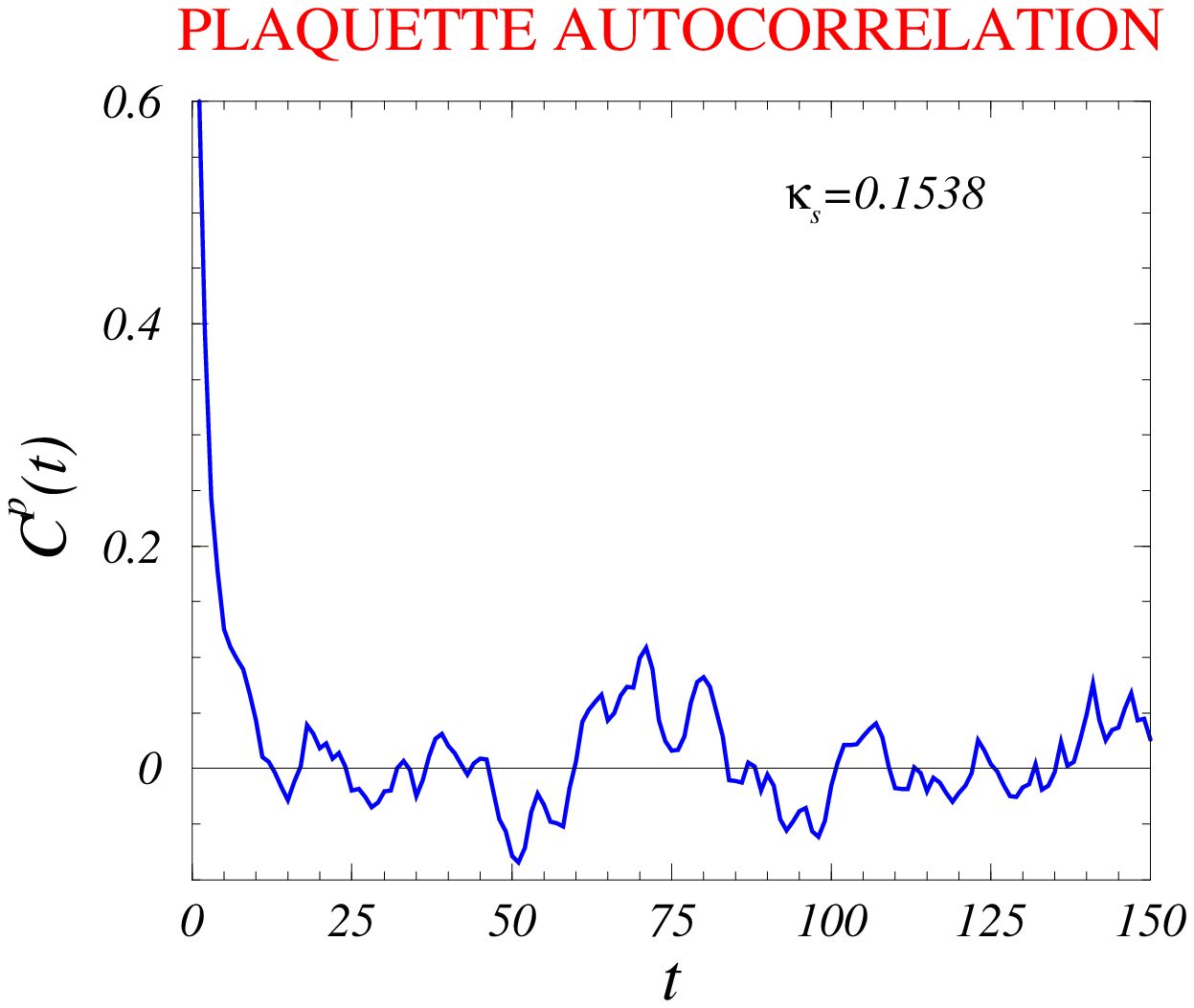}
\epsfxsize=8.0cm
\epsfbox{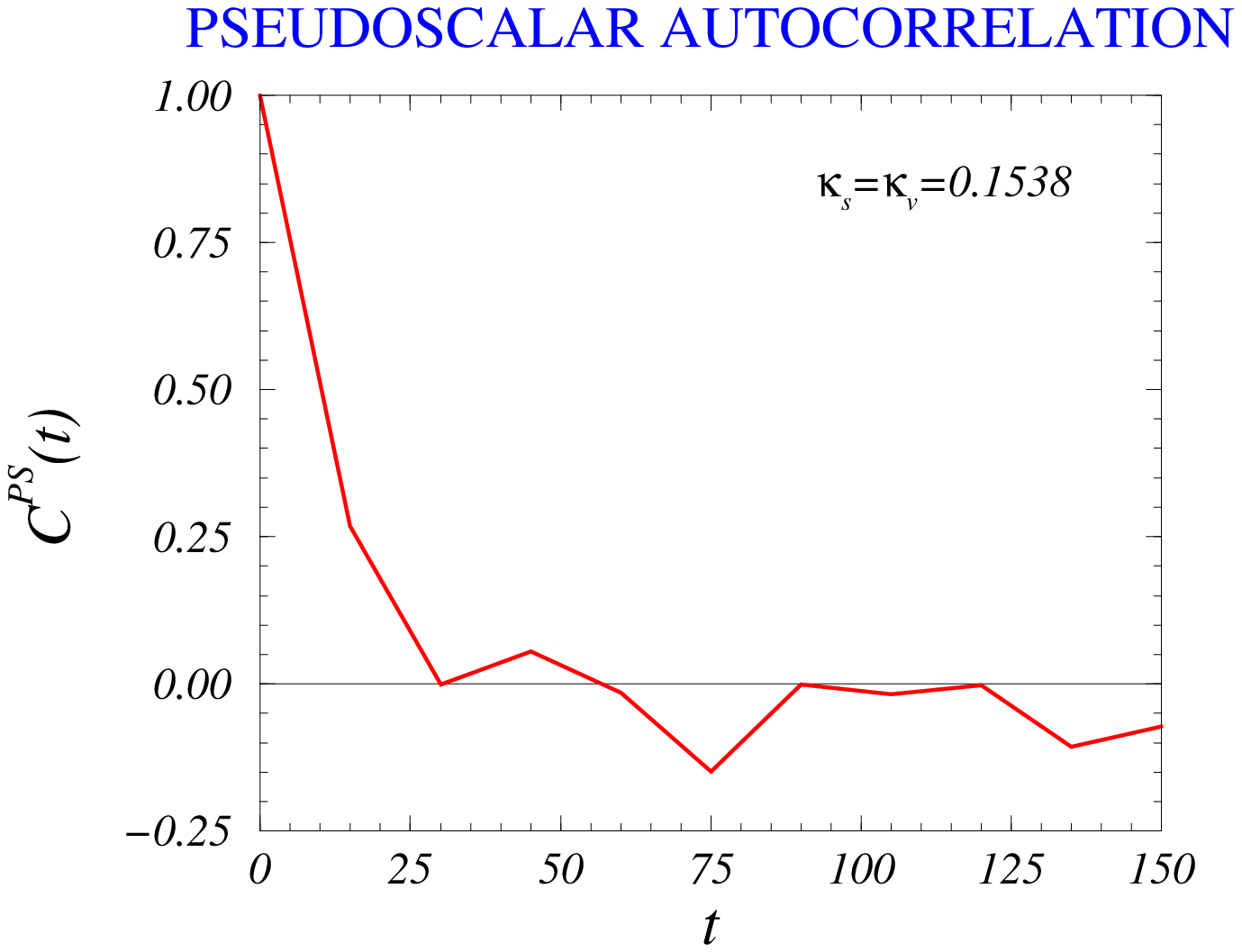}
\end{tabular}
\vspace*{-0.4cm}
\caption{\small \sl Autocorrelations of the plaquette $C^p(t)$ (left), and of 
the pseudoscalar correlation function $C^{PS}(t)$ (right), as a function
of the number of trajectories $t$ used to separate consecutive configurations.
For $C^{PS}(t)$ the fluctuations around zero start at
about $t = 30$, indicating that configurations separated by less than $30$
trajectories are correlated. The above plots refer to the run at $\beta=5.8$ 
($V=24^3 \times 48$, 
and $\kappa_s = 0.1538$).\label{fig:autocorr}}
\end{figure}

To select sufficiently decorrelated configurations we studied the
autocorrelation as a function of the number of trajectories $t$ by which 
two configurations are separated. 
For a given observable $A$, the autocorrelation is evaluated as
\be
 C^A(t)=\dfrac{\dfrac{1}{(T_{MC}-t)}\displaystyle{\sum_{s=1}^{T_{MC}-t}}
 (A_s-\langle A \rangle)(A_{s+t}-\langle A \rangle)}{\dfrac{1}{T_{MC}}
\displaystyle{ \sum_{s=1}^{T_{MC}}}(A_s-\langle A \rangle)^2}\,,
 \ee
where $\langle A \rangle$ is the mean value of $A$ computed on the total number 
of trajectories $T_{MC}$ and $A_s$ denotes the value measured along 
the trajectory $s$. 
After examining the autocorrelation of the plaquette and of 
the pseudoscalar two-point correlation functions, both shown in 
fig.~\ref{fig:autocorr}, we made the conservative choice to select  
configurations separated by $45$ trajectories.~\footnote{The autocorrelation of 
the pseudoscalar two-point function is made for the time separation between 
the sources fixed to $14$.}
The sea quark masses used in our simulations correspond to ratios of 
pseudoscalar over vector meson masses covering the range 
$0.60\lesssim M_P/M_V\lesssim 0.75$. 
In our data analysis we 
distinguish the sea ($\kappa_s$) from the valence ($\kappa_v$) hopping 
parameters, where, for each value of the sea quark mass, $\kappa_v$ can assume any
of the values listed in table~\ref{TTtable1}.
The main results of this paper refer to the run made at $\beta=5.8$ and $L=24$ 
(i.e., $L\simeq 1.5$~fm), whereas those obtained on the lattice with $L=16$ and 
$\beta$ either $5.8$ ($L\simeq 1$~fm) or $5.6$ ($L\simeq 1.3$~fm) are used to 
investigate finite volume and discretisation effects respectively.

Finally,  in order to assess the effects of quenching, we also made a
quenched 
run ($250$ configurations) at $\beta=6.2$ and $V=24^3\times 56$ ($L\simeq 1.6$ 
fm). 
To reproduce the values of the $N_f=2$ pseudoscalar meson masses,
we chose in the quenched run $\kappa_v\in\{0.1514,0.1517,0.1519,0.1524\}$.

An important feature of the data analysis in the partially quenched case  
is the appropriate treatment of the statistical errors. 
While the results of computations of an observable $O$ on configurations with 
different values of $\kappa_s$ are statistically independent, those computed on 
the gauge configurations with the same $\kappa_s$ but different $\kappa_v$ 
are not. 
To account for these correlation effects we use a combination of the
bootstrap 
and jackknife statistical methods. 
At fixed $\kappa_s$, our data are clustered into $N_{jk}$ subsets 
and the statistical errors on $O$ are estimated by using the jackknife method.   
The resulting $N_{jk}$ values at a given $\kappa_s$, $O_{jk}^{\kappa_s}$, 
are then considered as a set of uniformly distributed numbers. In a next step   
one generates $N_b\gg N_{jk}$ bootstrap events by taking, for each  $\kappa_s$, 
one of the uniformly distributed $O_{jk}^{\kappa_s}$. 
The variance of $O$ is then estimated by
\be
\label{eq:bootvar}
\sigma^2_O = (N_{jk}-1) \times \biggl(\langle O^2 \rangle_{N_b} - \langle O
\rangle_{N_b}^2\biggr)\,,
\ee
where $\langle ... \rangle_{N_b}$ stands for the average calculated over the
$N_b$ bootstrap events, while the factor $(N_{jk}-1)$ takes into account the
jackknife correlation at fixed $\kappa_s$. 
For a comparison, we also considered 
the procedure adopted in ref.~\cite{AliKhan}, 
consisting in applying the jackknife method at one specific value of $\kappa_s$,
while keeping the data at the remaining three $\kappa_s$ fixed at their central 
values. 
As expected, after adding in quadrature the four variances calculated in
this way, we find excellent agreement with the estimates obtained by
using eq.~(\ref{eq:bootvar}).

\begin{table}[!h]
\centering
\begin{tabular}{|c|c|c||c|c|c|}
\hline
$\kappa_{s}$ & $\kappa_{v_1}$ & $\kappa_{v_2}$ & $M_P$ & 
 $M_V$ & $m_{v12}^{\rm AWI}(a)$\\
\hline
$0.1535$ & $0.1535$ & $0.1535$ & $0.262(4)$ &$0.348(8)$ &$0.0333(6)$ \\
  & $0.1535$ & $0.1538$ & $0.252(5)$  &$0.341(8)$ &      $0.0306(6)$ \\
  & $0.1535$ & $0.1540$ & $0.245(5)$ &$0.336(9)$ &       $0.0289(6)$ \\
  & $0.1535$ & $0.1541$ & $0.241(5)$ &$0.334(9)$ &       $0.0280(6)$ \\
  & $0.1538$ & $0.1538$ & $0.241(5)$ &$0.333(9)$ &       $0.0280(6)$ \\
  & $0.1538$ & $0.1540$ & $0.233(5)$  &$0.329(9)$ &      $0.0263(6)$\\
  & $0.1538$ & $0.1541$ & $0.229(5)$&$0.326(9)$ &        $0.0254(6)$ \\
  & $0.1540$ & $0.1540$ & $0.226(5)$  &$0.324(9)$ &      $0.0246(6)$ \\
  & $0.1540$ & $0.1541$ & $0.222(5)$ &$0.322(10)$ &      $0.0237(6)$ \\
  & $0.1541$ & $0.1541$ & $0.218(5)$ &$0.319(10)$ &      $0.0228(6)$ \\
\hline
$0.1538$ & $0.1535$ & $0.1535$ & $0.258(3)$ & $0.335(4)$ &$0.0314(2)$ \\
 & $0.1535$ & $0.1538$ & $0.247(3)$ &$0.328(4)$ &         $0.0287(2)$ \\
  & $0.1535$ & $0.1540$ & $0.240(3)$ & $0.323(5)$ &       $0.0269(2)$\\
  & $0.1535$ & $0.1541$ & $0.236(4)$ & $0.321(5)$ &       $0.0261(2)$ \\
  & $0.1538$ & $0.1538$ & $0.236(4)$ & $0.321(5)$ &       $0.0260(2)$ \\
  & $0.1538$ & $0.1540$ & $0.228(4)$ & $0.316(6)$ &       $0.0243(2)$ \\
  & $0.1538$ & $0.1541$ & $0.224(4)$ &$0.314(6)$ &        $0.0234(2)$ \\
  & $0.1540$ & $0.1540$ & $0.220(4)$ &$0.311(6)$ &        $0.0226(2)$ \\
  & $0.1540$ & $0.1541$ & $0.216(5)$ & $0.309(7)$ &       $0.0217(2)$\\
  & $0.1541$ & $0.1541$ & $0.212(5)$ & $0.306(7)$ &       $0.0208(2)$\\
\hline
$0.1540$ & $0.1535$ & $0.1535$ & $0.258(2)$ & $0.345(5)$ &$0.0303(3)$\\
  & $0.1535$ & $0.1538$ & $0.247(2)$ & $0.338(6)$ &       $0.0277(3)$\\
  & $0.1535$ & $0.1540$ & $0.240(3)$ & $0.333(7)$ &       $0.0259(3)$\\
  & $0.1535$ & $0.1541$ & $0.236(3)$ &$0.331(7)$ &        $0.0250(3)$\\
  & $0.1538$ & $0.1538$ & $0.236(3)$ & $0.331(7)$ &       $0.0250(3)$ \\
  & $0.1538$ & $0.1540$ & $0.228(3)$ & $0.326(8)$ &       $0.0233(3)$\\
  & $0.1538$ & $0.1541$ & $0.225(3)$ & $0.324(8)$ &       $0.0224(3)$ \\
  & $0.1540$ & $0.1540$ & $0.221(3)$ & $0.321(9)$ &       $0.0215(4)$ \\
  & $0.1540$ & $0.1541$ & $0.217(3)$ & $0.319(9)$ &       $0.0207(4)$\\
  & $0.1541$ & $0.1541$ & $0.212(3)$ & $0.316(10)$ &      $0.0198(4)$\\
\hline
$0.1541$ & $0.1535$ & $0.1535$ & $0.234(6)$ & $0.318(10)$ &$0.0286(3)$ \\
  & $0.1535$ & $0.1538$ & $0.222(6)$ & $0.311(11)$ &       $0.0259(3)$\\
  & $0.1535$ & $0.1540$ & $0.213(6)$ & $0.307(11)$ &       $0.0242(3)$\\
  & $0.1535$ & $0.1541$ & $0.209(6)$ &$0.304(11)$ &        $0.0233(3)$\\
  & $0.1538$ & $0.1538$ & $0.209(6)$ & $0.304(11)$ &       $0.0233(3)$\\
  & $0.1538$ & $0.1540$ & $0.200(7)$ &$0.300(12)$ &        $0.0215(3)$\\
  & $0.1538$ & $0.1541$ & $0.196(7)$ & $0.298(12)$ &       $0.0206(3)$\\
  & $0.1540$ & $0.1540$ & $0.191(7)$ & $0.295(12)$ &       $0.0198(3)$\\
  & $0.1540$ & $0.1541$ & $0.187(7)$ & $0.293(13)$ &       $0.0189(3)$ \\
  & $0.1541$ & $0.1541$ & $0.182(7)$ & $0.291(13)$ &       $0.0180(4)$\\
\hline
\end{tabular}
\caption{\small \sl Pseudoscalar and vector meson masses, and the bare quark 
masses $m_{v12}^{\rm AWI}(a)=\frac{1}{2}\bigl[ m_{v1}(a)+ m_{v2}(a)\bigr]
^{\rm AWI}$
obtained by using the axial Ward identity method (cf. eq.~(\ref{AWI})). 
Results, in lattice units, refer to the simulation with $\beta=5.8$  and 
$V=24^3 \times 48$. 
Those obtained at $\beta=5.6$ are tabulated in appendix~1.\label{table2}}
\end{table}
\section{\label{sec:masses}Masses of light mesons and quarks}
In this section we present the results for the light pseudoscalar and vector 
meson masses, as well as the bare quark masses, obtained from our main data-set 
($\beta=5.8$, $L=24$). 
These quantities are extracted from the standard study of two-point
correlation functions, $C_{JJ}(t)$, in which we consider both degenerate and 
non-degenerate valence quarks, with valence quarks either equal or different 
from the sea quarks.
At large Euclidean time separations, $t\gg 0$, $C_{JJ}(t)$ is dominated by the 
lightest hadronic state and we fit it to the form:
\be
\label{eq:meff}
C_{JJ}(t) = \sum_{\vec x} \langle 0 \vert J({\vec x}, t) J^{\dagger}(0)
\vert 0 \rangle\, \stackrel{t\gg 0}{\longrightarrow}\,  
{Z_{JJ} \over 2 M_J} \left( e^{- M_J t} + \eta \,e^{- M_J (T - t)}
\right)\,,
\ee
where $\eta$ is the time reversal $(t \leftrightarrow T - t)$ symmetry 
factor.~\footnote{$\eta=+1$ for $C_{PP}(t)$, $C_{VV}(t)$, or $C_{AA}(t)$,  
whereas for $C_{AP}(t)$, $\eta=-1$. In this notation $P=\bar q \gamma_5 q$, 
$V\equiv V_i= \bar q \gamma_i q$, and 
$A\equiv A_0= \bar q \gamma_0 \gamma_5 q$.} 
The pseudoscalar and vector meson masses 
are extracted from $C_{PP}(t)$ and $C_{VV}(t)$ respectively, by fitting  
in $t\in [14,23]$. 
The corresponding results are listed in table~\ref{table2}. 
These results are obtained by using local source operators. 
We checked, however, that the Jacobi smearing procedure~\cite{Allton} 
leaves the masses unchanged, although the lowest hadronic state becomes
isolated earlier. 

In the same table~\ref{table2} we also give the results for the quark masses 
obtained by using the axial Ward identity (AWI) definition. 
More specifically, with  $A_0(x)=\bar q_{v1} \gamma_0 \gamma_5 q_{v2}$, and 
$P=\bar q_{v1} \gamma_5 q_{v2}$, 
\be
\label{AWI}
{\langle \displaystyle{\sum_{\vec x}} \partial_0
A_0(x) P^\dagger(0)\rangle \over  \langle \displaystyle{\sum_{\vec x}} P
(x) P^\dagger(0)\rangle } \,\stackrel{t\gg 0}{\longrightarrow}\,\bigl[ m_{v1}(a)
+ m_{v2}(a)\bigr]^{\rm AWI},
\ee
where we fit in the same interval as before, i.e., $t\in [14,23]$. 
Alternatively, one can use the vector Ward identity (VWI) and relate the bare 
quark mass to the Wilson
hopping parameter as:
\be\label{VWI}
m_v^{\rm VWI}(a) =  
\dfrac{1}{2}\left( \dfrac{1}{\kappa_v} - \dfrac{1}{\kappa_{cr}(\kappa_s)}
\right)\, ,
\ee
with $\kappa_v$ being either $\kappa_{v1}$ or $\kappa_{v2}$ 
(cf. table~\ref{table2}), and $\kappa_{cr} (\kappa_s)$ is the critical
hopping 
parameter for a given sea quark mass. 
Notice that both the $AWI$ and $VWI$ definitions of the quark mass do depend
on the values of the sea quark masses.
In the axial case, this dependence is provided by the ratio of two-point
correlation functions of eq.~(\ref{AWI}).
In the VWI definition, the dependence on the sea quark mass is given by
the value of the critical hopping parameter $\kappa_{cr}(\kappa_s)$
computed at
fixed $\kappa_s$.
We also notice that the sea quark dependence of the critical hopping parameter 
is numerically equivalent to the observation made in ref.~\cite{Gockeler} where 
such a dependence is expressed in terms of the renormalisation constants of the 
singlet and non-singlet scalar density.
 
\subsection{Lattice spacing and $\kappa_{cr}(\kappa_s)$}
\begin{figure}[t]
\centering
\begin{tabular}{cc}
\epsfxsize=8cm
\epsfbox{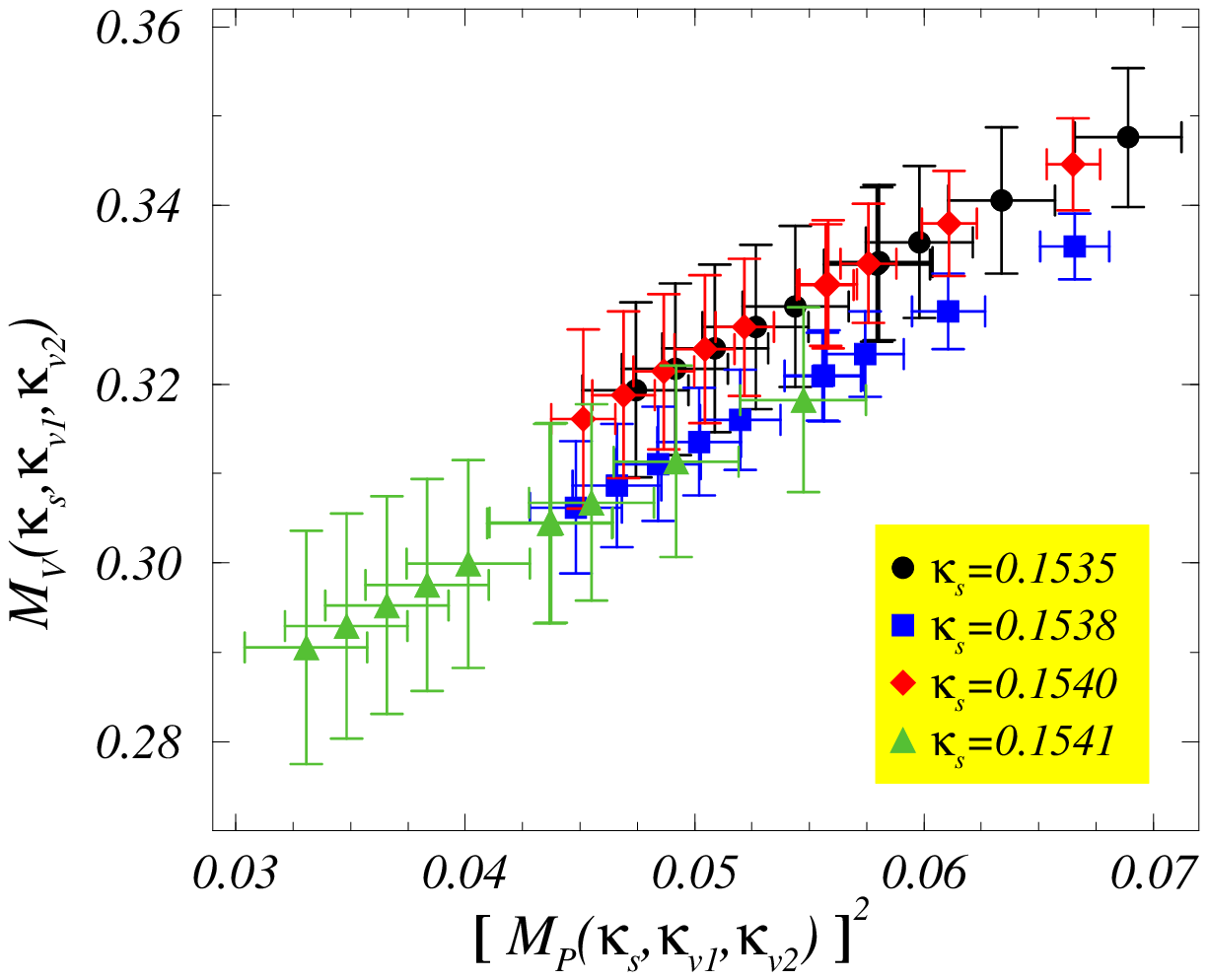}
\hspace*{0.1cm}
\epsfxsize=8cm
\epsfbox{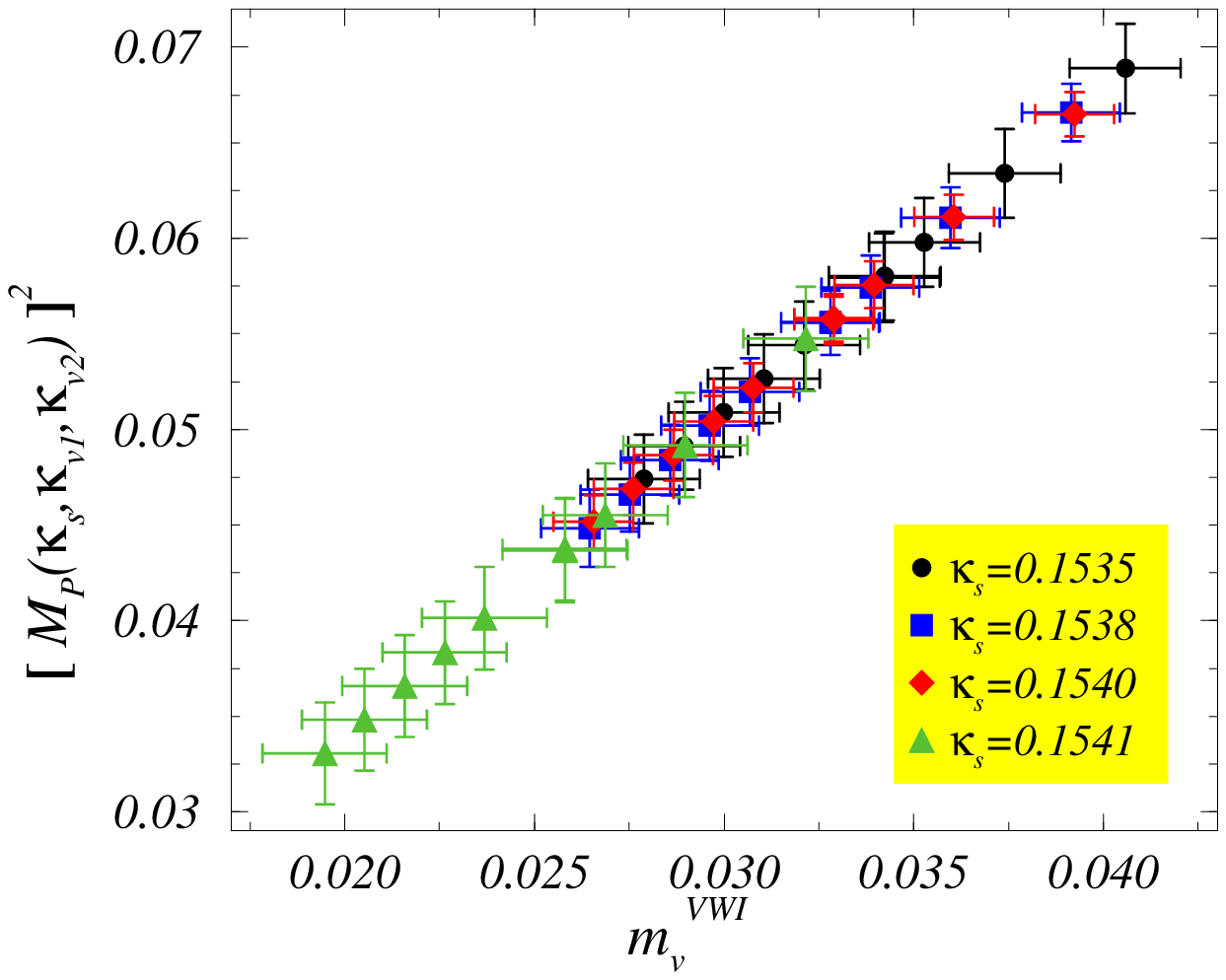}
\vspace*{-0.4cm}
\end{tabular}
\caption{\small \sl Data-points that illustrate the functional dependencies 
discussed in 
eqs.~(\ref{eq:mvvsmp}) and ~(\ref{eq:mp2vsk}).\label{fig:mvmp}}
\end{figure}

In order to fix the value of the lattice spacing and $\kappa_{cr}(\kappa_s)$, 
we inspect the dependence of the pseudoscalar and vector meson masses on the 
valence and sea quark masses.
The left panel of fig.~\ref{fig:mvmp} shows the dependence of the vector meson 
masses on the squared pseudoscalar ones, which suggests a simple linear fit to 
the form 
\be
\label{eq:mvvsmp}
M_V(\kappa_s,\kappa_{v_1},\kappa_{v_2}) = P_1 + P_2 \times M_P^2(\kappa_s,
\kappa_{v_1},\kappa_{v_2})  + P_3 \times  M_P^2(\kappa_s,\kappa_s,\kappa_s)\,,
\ee
in an obvious notation, and with capital letters used to denote that 
the meson masses are given in lattice units. 
Such a fit yields $P_3=0.1(3)$, i.e., the sea quark dependence is very weak. 
We can  therefore neglect it and set $P_3=0$. 
In this way we obtain $P_1=0.25(1)$ and $P_2=1.4(2)$. 
We then determine the lattice spacing from this fit and by using the method
of
the ``physical lattice planes"~\cite{Allton2}: at the intersection of our data 
with the physical value for the ratio $m_K/m_{K^\ast}=0.554$, we impose that 
$M_{K^\ast}$ obtained on the lattice is $M_{K^\ast} = a m_{K^{\ast}}^{phys}$,
thus obtaining 
\be
\label{eq:am1}
a^{-1}_{m_{K^\ast}}=3.2(1) \gev\,.
\ee
We also estimate the value of the lattice spacing by studying the static quark 
potential.
At each value of the sea quark mass we extract the following values of the 
Sommer parameter $R_0=r_0/a$~\cite{sommer}:
\be 
\left(R_0\right)_{\kappa_s}= \{7.52(17)_{0.1535},\,7.70(9)_{0.1538},\,
7.78(19)_{0.1540},\,8.10(20)_{0.1541}\}. 
\ee
Details on this extraction can be found in appendix~2. 
Assuming the observed dependence of $R_0$ on the sea quark mass to be physical, 
as the lattice spacing $a$ is supposed to be independent on the sea quark mass, 
we linearly extrapolate the above results to the chiral limit 
[$R_0 = \alpha +\beta \times M_P^2(\kappa_s,\kappa_s,\kappa_s)$], 
ending up with $R_0=8.6(4)$. 
After setting $r_0=0.5$~fm, we find
\be
\label{eq:am2}
a^{-1}_{r_0}=3.4(2) \gev \,,
\ee
in good agreement with the value given in eq.~(\ref{eq:am1}).

Finally, for $m_v^{\rm VWI}$ in eq.~(\ref{VWI}) we also need 
$\kappa_{cr}(\kappa_s)$. 
Inspired by partially quenched ChPT~\cite{sharpe}, but neglecting the chiral 
logs since our dynamical quarks are not light enough, we fit our data to 
\be
\label{eq:mp2vsk}
 M_P^2(\kappa_s,\kappa_{v_1},\kappa_{v_2}) = Q_1 \, m_v^{VWI} 
 [1  + Q_2 \,  m_s^{VWI} + Q_3 \,  m_v^{VWI}]\,,
\ee
and obtain $Q_2=-0.2(25)$ and $Q_3=0.4(14)$. 
In other words our data are not sensitive to the quadratic 
corrections proportional to $Q_2$ and $Q_3$, as it can be seen from the right 
plot in fig.~\ref{fig:mvmp}.  
Very similar feature is observed if instead of $m_{v,s}^{VWI}$ we use 
$m_{v,s}^{AWI}$, and therefore in the following we set $Q_2=Q_3=0$. 
From the resulting fit to eq.~(\ref{eq:mp2vsk}), we get $Q_1=1.70(3)$ and 
\bea
(\kappa_{cr})_{\kappa_s}= \left\{0.15544(7)_{0.1535},0.15537(6)_{0.1538},
0.15537(5)_{0.1540},0.15503(8)_{0.1541}\right\}\,.
\eea

\section{\label{sec:NPR}Quark mass renormalisation}
Before discussing the strategy to identify the physical strange and the average 
up/down quark masses from the results obtained by using 
eqs.~(\ref{AWI},\ref{VWI}), we need to determine the corresponding 
multiplicative mass renormalisation constants, 
$Z_{m}^{AWI}(a\mu)=Z_A/Z_P(a\mu)$, and $Z_{m}^{VWI}(a\mu)=1/Z_S(a\mu)$. 
For completeness, we also present in this section the results for $Z_V$, $Z_T$
and the quark field RC $Z_q$.

We use the $\ri$ method~\cite{Martinelli} which we explained in great
detail in a previous publication~\cite{Becirevic}. We adopt the same
notation as in 
ref.~\cite{Becirevic} and compute the renormalisation group invariant 
combinations
\be
\label{eq:zmu0}
Z_{\mathcal{O}}(a\mu_0) = C_{\mathcal{O}}(a\mu_0) \, Z_{\mathcal{O}}^{RGI} =
C_{\mathcal{O}}(a\mu_0) \, \left( Z_{\mathcal{O}}(a\mu)/C_{\mathcal {O}}
(a\mu)\right) \,,
\ee
for the scale dependent bilinear operators, $\mathcal{O}=S, P, T$. 
The evolution functions $C_{\mathcal{O}}(a\mu)$, which are known in 
the $\ri$ scheme at the ${\rm N^3LO}$ for $Z_S$ and $Z_P$~\cite{Chetyrkin} and
at the ${\rm N^2LO}$ for $Z_T$~\cite{Gracey:2003yr}, 
explicitly cancel the scale dependence of the RCs, 
$Z_{\mathcal{O}}(a\mu)$, at the considered order in continuum perturbation
theory. 

\begin{figure}[h]
\centering
\epsfxsize=8cm
\epsfbox{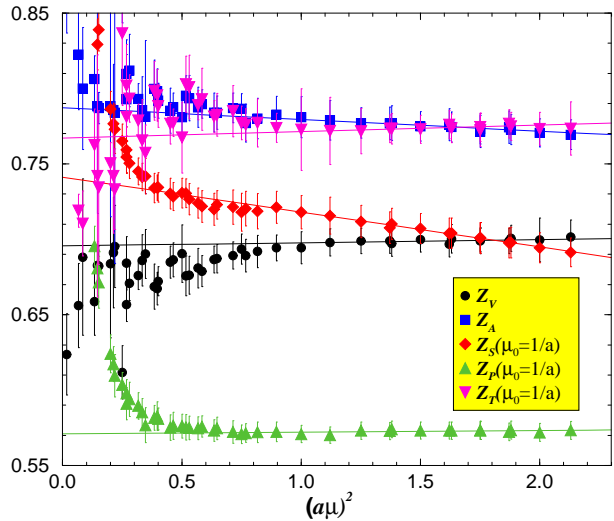}
\vspace*{-0.4cm}
\caption{\small \sl Renormalisation constants in the $\ri$ scheme at the
scale $a\mu_0=1$, computed according to eq.~(\ref{eq:zmu0}), as a function
of the initial renormalisation scale $(a \mu)^2$. Discretisation effects 
$\propto (a \mu)^2$ are eliminated by extrapolating the data from 
$1\leq (a\mu)^2\leq 2$, to $(a \mu)^2=0$. Illustration is provided for the
data with $\kappa_s=0.1538$.\label{fig:zlinear}}
\end{figure}

In fig.~\ref{fig:zlinear} we plot  $Z_{V,A}$, as well as $Z_{S,P,T}(a\mu_0)$, 
where in  eq.~(\ref{eq:zmu0}) we choose $a\mu_0=1$.  
All of them are supposed to be flat in $(a\mu)^2$, where $\mu$ is the
initial renormalisation scale introduced in eq.~(\ref{eq:zmu0}),
modulo discretisation effects. 
In fact, however, we see that these effects are pronounced in the case of 
$Z_S(a\mu_0)$, which we correct for by linearly extrapolating from 
$1\leq (a\mu)^2\leq 2$, to $(a\mu)^2=0$.
Furthermore, for the $\ri$ scheme to be mass independent we extrapolate 
the mild sea quark dependence linearly to the chiral limit. The results 
are reported in table~\ref{tab:renRI}. 
By confronting the results for the RCs obtained at $\beta=5.8$ and $L=24$
with those obtained at the same $\beta$ but with $L=16$, we find that they
differ by less than $5$\%. 
Given the smallness of the physical lattice size at $L=16$ ($\sim 1$ fm),
we expect finite volume effects on our estimate of the RCs to be
negligible with respect to discretization errors, which are evaluated to
be of the order of $10$\% (see subsec.~\ref{subsec:systematics}).
This is expected, since the RCs encode the short distance physics while
the volume finiteness affects the long distance physics.
\begin{table}[h!]
\centering
\begin{tabular}{|cc|cccccc|}
\hline
{\phantom{\huge{l}}} \raisebox{-.1cm} {\phantom{\huge{j}}}
$\beta$ ($N_f$) & $\kappa_s$ & $Z_q$ & $Z_V$ & $Z_A$  & $Z_S$ & $Z_P$ & $Z_T$ \\
\hline\hline
{\phantom{\Large{l}}} \raisebox{-.1cm} {\phantom{\Large{j}}}
$5.8$ ($N_f=2$) & $0.1535$   & $0.81(1)$ & $0.70(1)$ &$0.80(2)$ & $0.74(1)$ & 
                $0.54(1)$ & $0.78(2)$\\
{\phantom{\Large{l}}} \raisebox{-.1cm} {\phantom{\Large{j}}}
                & $0.1538$   & $0.80(1)$ & $0.70(1)$ & $0.79(2)$ & $0.73(2)$ &
		$0.55(1)$ & $0.77(2)$ \\
{\phantom{\Large{l}}} \raisebox{-.1cm} {\phantom{\Large{j}}}
                & $0.1540$   & $0.78(1)$ & $0.67(1)$ & $0.77(2)$ & $0.72(1)$ &
		$0.52(1)$ & $0.75(2)$ \\
{\phantom{\Large{l}}} \raisebox{-.1cm} {\phantom{\Large{j}}}
                & $0.1541$   & $0.79(1)$ & $0.69(1)$ & $0.78(2)$ & $0.71(2)$ &
		$0.52(1)$ & $0.77(2)$ \\
{\phantom{\Large{l}}} \raisebox{-.1cm} {\phantom{\Large{j}}}
 &$\kappa_{cr}(\kappa_{cr})$ & $0.76(1)$ & $0.66(2)$ & $0.76(1)$ & $0.67(3)$ &
 $0.50(2)$ & $0.74(4)$ \\
\hline
{\phantom{\Large{l}}} \raisebox{-.1cm} {\phantom{\Large{j}}}
  & N-BPT & 0.75 & $0.70$ & $0.77$ & $0.75$ & $0.61$ & $0.75$\\
{\phantom{\Large{l}}} \raisebox{-.1cm} {\phantom{\Large{j}}}
  & TI-BPT & 0.72 & $0.65$ & $0.73$ & $0.71$ & $0.55$ & $0.71$\\
\hline\hline
{\phantom{\Large{l}}} \raisebox{-.1cm} {\phantom{\Large{j}}}
$5.6$ ($N_f=2$) & $0.1560$   & $0.78(2)$ & $0.63(4)$ & $0.78(3)$ & $0.67(2)$ & $0.46(1)$ & $0.76(5)$\\
{\phantom{\Large{l}}} \raisebox{-.1cm} {\phantom{\Large{j}}}
                & $0.1575$   & $0.78(2)$ & $0.64(4)$ & $0.77(2)$ & $0.65(2)$ &
		$0.46(1)$ & $0.77(3)$\\
{\phantom{\Large{l}}} \raisebox{-.1cm} {\phantom{\Large{j}}}
                & $0.1580$   & $0.78(1)$ & $0.66(1)$ & $0.78(4)$ & $0.69(2)$ &
		$0.50(1)$ & $0.78(3)$\\
{\phantom{\Large{l}}} \raisebox{-.1cm} {\phantom{\Large{j}}}
&$\kappa_{cr}(\kappa_{cr})$ & $0.78(1)$ & $0.66(1)$ & $0.78(1)$ & $0.67(1)$ &
$0.47(1)$ & $0.78(2)$\\
\hline
{\phantom{\Large{l}}} \raisebox{-.1cm} {\phantom{\Large{j}}}
  & N-BPT & 0.74 & $0.67$ & $0.75$ & $0.73$ & $0.58$ & $0.73$\\
{\phantom{\Large{l}}} \raisebox{-.1cm} {\phantom{\Large{j}}}
  & TI-BPT & 0.69 & $0.62$ & $0.71$ & $0.69$ & $0.51$ & $0.68$\\
\hline\hline
{\phantom{\Large{l}}} \raisebox{-.1cm} {\phantom{\Large{j}}}
$6.2$ ($N_f=0$) & $-$ &$0.82(1)$ &$0.71(1)$ & $0.80(1)$ & $0.71(1)$ & $0.54(1)$
& $0.79(3)$\\
\hline
{\phantom{\Large{l}}} \raisebox{-.1cm} {\phantom{\Large{j}}}
  & N-BPT & 0.78 & $0.73$ & $0.79$ & $0.77$ & $0.65$ & $0.77$\\
{\phantom{\Large{l}}} \raisebox{-.1cm} {\phantom{\Large{j}}}
  & TI-BPT & 0.72 & $0.64$ & $0.73$ & $0.71$ & $0.55$ & $0.71$\\
\hline
\end{tabular}
\caption{\small \sl Values of the renormalisation constants in the $\ri$
scheme at the scale $\mu = 1/a$ for the two partially quenched simulations
at $\beta=5.8$ ($a^{-1}=3.2(1)\gev$) and 5.6 ($a^{-1}=2.4(2)\gev$) and for the
quenched simulation at $\beta=6.2$ ($a^{-1}=3.0(1)\gev$). The quoted errors
are statistical only.
The non-perturbatively determined RCs are confronted to their perturbative 
value (BPT), where ``TI" and ``N" stand for the two types of boosting the bare 
lattice coupling as discussed in the text. 
Definition of $Z_q$ is the same one as in eq.~(5) of ref.~\cite{Becirevic}. 
\label{tab:renRI}}
\end{table}
For an easier comparison of the non-perturbatively and perturbatively estimated 
RCs, in table~\ref{tab:renRI} we also give the RCs evaluated by means of 
$1$-loop BPT~\cite{Lepage}. 
These are obtained by replacing in the one-loop expressions of 
ref.~\cite{Capitani} the bare lattice coupling by a boosted one, 
$g_0^2\to \widetilde g^2$. 
We distinguish the so-called ``na\"\i ve" boosting 
\be
\tilde g^2 ={g_0^2}/\langle P\rangle\,,
\label{eq:gnBPT}
\ee
with $\langle P\rangle\ $ being the average plaquette, and the so called 
``tadpole improved" coupling, which 
is defined by inverting the perturbative series of the logarithm of the 
plaquette, namely~\footnote{For a more extensive discussion about various
forms of boosting, see ref.~\cite{Crisafulli}.}
\be
\ln{\langle P \rangle} = - \dfrac{1}{3}\,
\tilde g^2(3.40/a)\, \left[1 - (1.1905-0.2266 N_f)\,
\frac{\tilde g^2(3.40/a)}{(4\pi)}\right]\,.
\label{eq:gtiBPT}
\ee
The corresponding results for the RCs are then referred to as N-BPT and
TI-BPT respectively. 
From the results in table~\ref{tab:renRI} we see that the perturbative
estimates
of $Z_{S,P}$ are larger than the non-perturbative ones. 
In other words the non-perturbatively computed $Z_m^{AWI}=Z_A/Z_P$ 
($Z_m^{VWI}=1/Z_S$), at $\beta=5.8$, is larger by about $15$\% ($6$\%) than its 
BPT counterpart. 
These differences directly propagate to the lattice determination of the 
renormalised quark masses. 
As we shall see in the following this effect explains the differences 
between our non-perturbatively renormalised results and those obtained by the 
CP-PACS and JLQCD Collaborations~\cite{AliKhan,Aoki}, where the TI-BPT has been 
used. 
Furthermore, our results agree with those obtained in 
refs.~\cite{Gockeler,DellaMorte}, in which the non-perturbative renormalisation 
has been implemented.

\section{\label{sec:result}Physical light quark masses}

\begin{figure}[t]
\centering
\begin{tabular}{cc}
\epsfxsize=8cm
\epsfbox{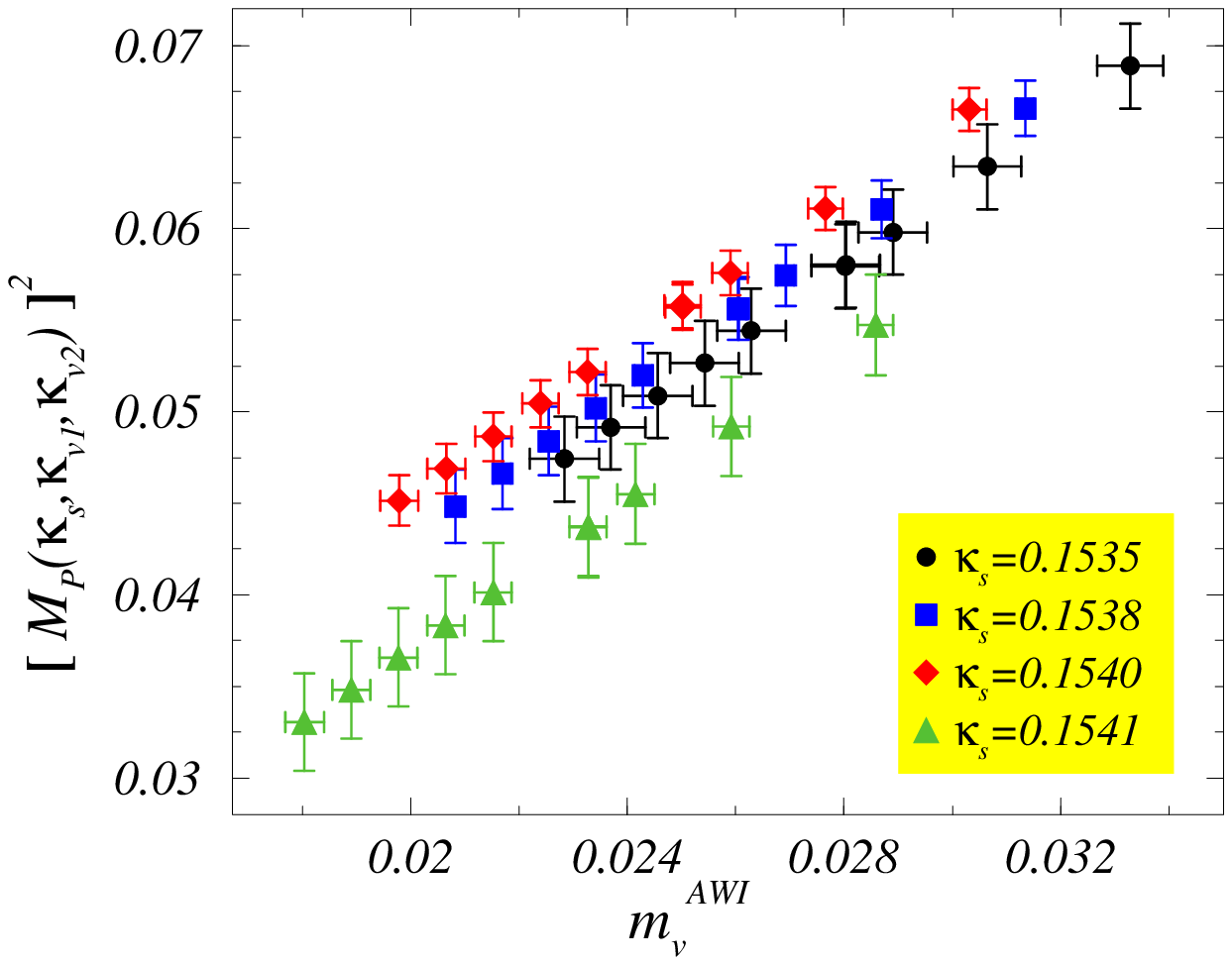}
\epsfxsize=8cm
\epsfbox{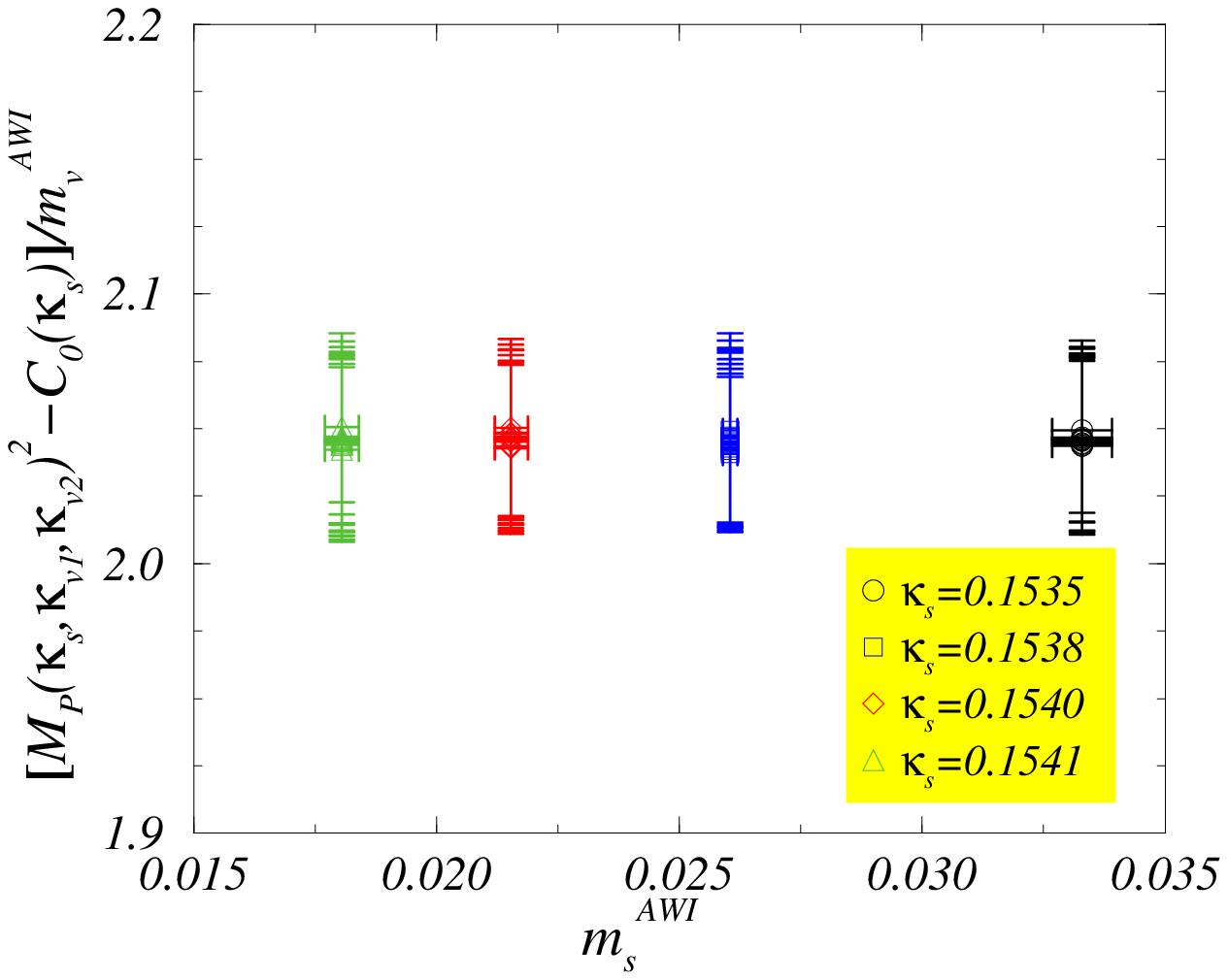}
\vspace*{-0.4cm}
\end{tabular}
\caption{\small \sl Dependence of the pseudoscalar meson masses on the
valence quark masses $m_v^{AWI}$ (left) and of the ratios
$(M_P^2-C_0)/m_v^{AWI}$ (cf. eq.~(\ref{eq:mp2vsaro})) on
the sea quark masses $m_s^{AWI}$. The results refer to the case of the
partially quenched simulations at $\beta=5.8$.\label{fig:mp2vsaro}}
\end{figure}

In order to determine the physical values of the light quark masses, we
need to investigate the dependence of the pseudoscalar meson masses
squared on the valence and sea quark masses. From the left plot in
fig.~\ref{fig:mp2vsaro} we see that: (i) the dependence on
the valence quark mass is linear, (ii) the slopes do not depend on
$\kappa_s$, and (iii) the intercepts do depend on $\kappa_s$ and they are
non-zero even when the sea quark mass is sent to zero. 
We can therefore fit to the form
\be
\label{eq:mp2vsaro}
M_P^2(\kappa_s,\kappa_{v_1},\kappa_{v_2}) =  C_0(\kappa_s) + C_1 \,
 m_v^{AWI} [1 + C_2 \, m_s^{AWI} + C_3 \, m_v^{AWI}]\,,
\ee
where the valence and sea AWI quark masses are defined in eq.~(\ref{AWI}). 
In this notation $m_v^{AWI} \equiv m^{AWI}(\kappa_s,\kappa_{v_1},\kappa_{v_2})$,
and $m_s^{AWI} \equiv m^{AWI}(\kappa_s,\kappa_s,\kappa_s)$.
Obviously, the fit forms~(\ref{eq:mp2vsk}) and~(\ref{eq:mp2vsaro}) differ
not only in the use of differently defined quark masses (AWI vs. VWI), but
also in 
the presence of the constant term $C_0(\kappa_s)$ in
eq.~(\ref{eq:mp2vsaro}). 
This term accounts for the ${\cal O}(a)$-discretisation
effects present in either the pseudoscalar meson masses or the AWI quark
masses or in both. Clearly, discretisation effects in the pseudoscalar
meson masses also affect the fit to eq.~(\ref{eq:mp2vsk}), i.e., the
determination of the critical hopping parameters $\kappa_{cr}(\kappa_s)$.

As before, we find that the values of the coefficients of the
quadratic terms, $C_2$ and $C_3$, in eq.~(\ref{eq:mp2vsaro}) are consistent
with zero [$C_2=0.6(18)$, and $C_3=0.08(63)$], which can be also seen from 
the right plot in fig.~\ref{fig:mp2vsaro}. Therefore we may set $C_2=C_3=0$, 
and from the fit to eq.~(\ref{eq:mp2vsaro}) we obtain  
$C_1=2.05(3)$, and 
\bea
C_0(\kappa_s)=\left\{0.0007(18)_{0.1535},\,0.0023(20)_{0.1538},\,
0.0046(15)_{0.1540},\,-0.0039(25)_{0.1541}
\right\}\,.
\eea

Our partially quenched estimates of the average up/down ($m_{ud}$) and of the 
strange ($m_s$) quark masses are made by substituting on the l.h.s. of 
eqs.~(\ref{eq:mp2vsk}) and~(\ref{eq:mp2vsaro}) the physical pion and kaon masses
and using the lattice spacing value given in either eq.~(\ref{eq:am1}) or 
eq.~(\ref{eq:am2}). 
In other words, to get the quark masses in physical units, we solve 
\bea
&& \left( m_\pi^2\right)^{phys.} = Q_1\times a^{-1}\times  m_{ud} \,,\nn \\
&& \left( m_K^2\right)^{phys.}= \frac{Q_1}{2} \times a^{-1}\times
\left( m_s +  m_{ud}\right)\,,
\eea
in the VWI case, and the analogous expressions, with the coefficient $Q_1$
substituted by $C_1$, in the AWI case. Finally the bare quark masses are 
multiplied by the corresponding $\ri$ mass renormalisation constants
computed at $\mu_0 = 1/a$, already listed in table~\ref{tab:renRI}. 

The conversion to the reference $\msb$ scheme, at $\mu=2\gev$, 
is made by using the perturbative expressions known up to N$^3$LO
accuracy~\cite{Chetyrkin,vermas}, and $\Lambda_{\msb}^{(N_f=2)}=
0.245(16)(20)\mev$~\cite{lambda}. We finally obtain 
\bea\label{res0}
{\rm VWI} &:&m_{ud}^{\msb}(2\gev) = 5.1(4) \mev \,, \quad m_s^{\msb}(2\gev) = 
120(7)\mev\,,\nn\\
{\rm AWI} &:&m_{ud}^{\msb}(2\gev) = 4.3(4) \mev\,, \quad m_s^{\msb}(2\gev) = 
101(8)\mev\,.
\eea
Notice that our VWI result is in very good agreement with the value 
$m_s^{\msb}(2\gev) = 119(5)(8)\mev$ obtained by the QCDSF-UKQCD 
collaboration who uses the VWI quark mass definition and the $\ri$ 
non-perturbative renormalisation~\cite{Gockeler}. 
Notice also that our AWI result agrees with the value obtained by the
Alpha collaboration, $m_s^{\msb}(2\gev) = 97(22)\mev$~\cite{DellaMorte}.

\subsection{Systematic uncertainties}
\label{subsec:systematics}
\begin{itemize}
\item[$\circ$] {\underline{\it Discretisation I}}:
From quenched studies we learned that a discrepancy between the quark mass
values obtained by using two {\it a priori} equivalent definitions, namely
the VWI and AWI definitions, is due to discretisation errors. Such a
difference has also been observed in other lattice studies with
$N_f=2$~\cite{AliKhan,Aoki}. In particular, in ref.~\cite{AliKhan} 
this difference is shown to diminish as the lattice spacing is reduced
so that, in the continuum limit, the quark masses defined in the two ways
converge to the same (unique) value. From that study it became clear that
discretisation errors are significantly larger in the VWI determination of
the quark masses which is why the JLQCD collaboration choose to quote
their AWI masses as final results, while the difference between the VWI
and AWI results is included in the asymmetric systematic error. We
follow the same procedure here and we obtain
\bea
\label{eq:res}
m_{ud}^{\msb}(2\gev)=4.3(4)(^{+0.8}_{-0})\mev\,, &
m_s^{\msb}(2\gev)=101(8)(^{+19}_{-0})\mev\,.
\eea

\item[$\circ$] {\underline{\it Discretisation II}}: 
The results reported in eq.~(\ref{res0}) refer to the lattice with
coupling $\beta=5.8$ (and volume $24^3\times 48$). From our data produced
at $\beta=5.6$ (see table~\ref{TTtable1}) we obtain
\bea
\label{eq:res56}
m_{ud}^{\msb}(2\gev)=4.2(3)(^{+0.7}_{-0})\mev\,, & 
m_s^{\msb}(2\gev)=101(6)(^{+18}_{-0})\mev\,,
\eea
thus indistinguishable from those obtained at $\beta=5.8$. 
This makes us more confident that the discretisation error attributed to
our results in eq.~(\ref{eq:res}) is realistic. 

\item[$\circ$] {\underline{\it Lattice spacing}}: 
To obtain the results quoted in eq.~(\ref{eq:res}) we used the lattice 
spacing~(\ref{eq:am1}) fixed by $m_{K^\ast}^{phys.}$. 
These results would become larger by $6$\% if we used 
$a^{-1}_{r_0}$~(\ref{eq:am2}). 
We will add this difference to our final error budget.

\item[$\circ$] {\underline{\it Renormalisation constants}}: 
As discussed in the text the determination of the mass renormalisation
factors is sensitive to discretisation errors. They are especially
important when the unimproved Wilson quark action is used, which is the
case with our study. To check for these effects we determined $Z_P/Z_S$ by
using the hadronic Ward identities (hWI)~\cite{bochicchio} (see
eqs.~(16,17) of ref.~\cite{Becirevic}), obtaining
\bea
\left(Z_P/Z_S\right)^{\rm hWI} = 0.67(1) \,.
\eea
Compared to the results in table~\ref{tab:renRI}, where at $\beta=5.8$ we
find $(Z_P/Z_S)^{\rm RI-MOM}=0.75(1)$, this result is about $11$\%
smaller. A similar effect is seen at $\beta=5.6$. We observe that, either
one attributes this effect to $Z_P$ or to $Z_S$, $(Z_P/Z_S)^{\rm hWI}$
always goes in the direction of decreasing the AWI-VWI mass difference.
Therefore, this $11$\% uncertainty, being already enclosed in the $19$\% 
systematic error due to the AWI-VWI mass difference, will not be
added. We finally notice that the result obtained for
$Z_P/Z_S$ by using the numerical stochastic perturbation theory to
$4$-loops, i.e., $(Z_P/Z_S)^{\rm NSPT}=0.77(1)$~\cite{direnzo}, is more
consistent with our larger NPR-ed (i.e., $\ri$) value. 

\item[$\circ$] {\underline{\it Finite volume}}: 
Our data at $\beta=5.8$ and $16^3\times 48$ (see appendix 1) indicate
large finite volume effects, when compared to our main data-set with
$\beta=5.8$ and $24^3\times 48$. 
As an illustration we show in fig.~\ref{fig7} the effective mass 
plots obtained in the two volumes at the same value of the sea quark
mass.
\begin{figure}[h!]
\centering
\begin{tabular}{c}
\epsfxsize=8cm
\epsfbox{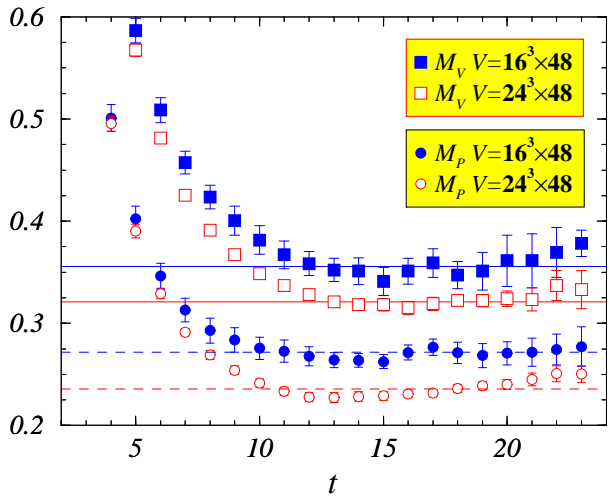}
\vspace*{-0.4cm}
\end{tabular}
\caption{\small \sl Pseudoscalar and vector effective mass plots at two 
different volumes at $\beta=5.8$ and with $\kappa_s=0.1538$. 
Valence quark masses are equal to that of the sea quark. 
The physical volumes correspond to $(1.5~{\rm fm})^3$ and $(1~{\rm fm})^3$ 
respectively.\label{fig7}}
\end{figure}
Such a sensitivity to the finiteness of the lattice box has already been 
observed in ref.~\cite{orth}, where it was shown that, for the range of
sea quark masses used in our study, finite volume effects are large at
$L\simeq 1$~fm but become insignificant when working on lattices with
$L\geq 1.5$~fm. 
We will rely on that conclusion and will assume that finite volume effects
are negligible with respect to our statistical and other sources of
systematic errors. For a more refined estimate of these effects
calculations on lattices with larger volumes are required.
\end{itemize}
We add the above systematic errors algebraically and, as a final result,
we quote
\be
\label{eq:FINAL}
\renewcommand{\arraystretch}{1.5}
\begin{array}{ll}
m_{ud}^{\msb}(2\gev)=4.3\pm 0.4^{+1.1}_{-0}~\mev\, \\
m_s^{\msb}(2\gev)=101\pm 8^{+25}_{-0}~\mev\,
\end{array}\qquad
\left(\begin{array}{cc} N_f=2 \,, \quad \beta=5.8\\ \ri\end{array}\right)\,,
\renewcommand{\arraystretch}{1.0}
\ee
which are the values already given in the abstract and the introduction
of the present paper.

\subsection{Impact of non-perturbative renormalisation and quenching
effects}
Before concluding this section we should compare our main results for the
light quark masses quoted in eq.~(\ref{eq:FINAL}) with those obtained by
evaluating the quark mass RCs using one-loop tadpole improved BPT:
\be
\label{eq:tiBPTqm}
\renewcommand{\arraystretch}{1.5}
\begin{array}{ll}
m_{ud}^{\msb}(2\gev)=3.7\pm 0.3^{+1.3}_{-0}~\mev\, \\
m_s^{\msb}(2\gev)=88\pm 7^{+30}_{-0}~\mev\,
\end{array}\qquad
\left(\begin{array}{cc} N_f=2 \,, \quad \beta=5.8\\
\mbox{TI-BPT}\end{array}\right)\,.
\renewcommand{\arraystretch}{1.0}
\ee
These results are in good agreement with those available in the literature
obtained from lattice stu\-di\-es with $N_f=2$ in which renormalisation 
is implemented perturbatively, namely  
$m_s^{\msb}(2\gev) = 88(^{+4}_{-6})\mev$~\cite{AliKhan}, and 
 $m_s^{\msb}(2\gev) = 84.5(^{+12.0}_{-1.7})$~MeV~\cite{Aoki}. 
Therefore, with Wilson-like dynamical fermions, the non-perturbative
renormalisation increases the values of the quark masses.

Finally, when compared with our results obtained in the quenched
simulation at 
$\beta=6.2$, 
\be
\label{eq:res_que}
\renewcommand{\arraystretch}{1.5}
\begin{array}{ll}
m_{ud}^{\msb}(2\gev)=4.6\pm 0.2^{+0.5}_{-0}~\mev\, \\
m_s^{\msb}(2\gev)=106\pm 2^{+12}_{-0}~\mev\,
\end{array}\qquad
\left(\begin{array}{cc} N_f=0 \,, \quad \beta=6.2\\ \ri\end{array}\right)\,,
\renewcommand{\arraystretch}{1.0}
\ee
on a lattice similar in size and resolution and by using the same
(unimproved) Wilson quark action, we see no evidence for any effect that
could be attributed to the presence of the dynamical quarks. For a clearer
conclusion concerning this point one should work with lighter sea quark
masses which is one of the main goals of our future simulations.

\section{\label{sec:concl}Conclusions}
In this paper we have presented our results for the light quark masses
obtained from a numerical simulation of QCD on the lattice with $N_f=2$
degenerate dynamical 
quarks, with masses covering the range $3/4\lesssim m_q/m_s^{phys}\lesssim
3/2$ 
(with respect to the physical strange quark mass). 
Our main values for the strange and for the average up/down quark masses 
are obtained on a lattice with spacing $a\approx 0.06$~fm and spatial 
volume $(1.5~{\rm fm})^3$, and by using the axial Ward identity. 
An important feature of our calculation is that we renormalise the quark masses 
non-perturbatively in the RI-MOM scheme, which is then converted to the $\msb$ 
scheme by using the known $4$-loop perturbative formulae. 
We show that 
\begin{itemize}
\item[--] within our statistical accuracy and with relatively heavy dynamical 
quarks, our unquenched results are fully consistent with the quenched 
ones;~\footnote{Both with the quenched results presented in 
this paper, and with those our collaboration presented before~\cite{ours}.}
\item[--] the non-perturbative renormalisation leads to resulting quark 
masses larger than those renormalised by using $1$-loop perturbation theory 
(even if tadpole improved);
\item[--] the systematic errors are dominated by lattice discretisation 
artifacts, which can be cured by implementing $\mathcal{O}(a)$-improvement
and/or working with several small lattice spacings followed by an extrapolation 
to the continuum limit.
\end{itemize}
The finiteness of the lattice volume is expected not to be an important source 
of systematic errors for the sea quark masses explored in our simulation. 
Further decrease of the dynamical quark masses would necessitate, however,
to work with larger physical lattice volumes. 
Our final results are given in eq.~(\ref{eq:FINAL}).

\section*{Acknowledgements}
We warmly thank G.~Martinelli for many useful discussions and
F.~Di Renzo for communicating to us their high precision perturbative result  
for $Z_P/Z_S$  prior to publication. 
The work by V.G. has been funded by MCyT, Plan Nacional I+D+I (Spain) under the 
Grant BFM2002-00568, and the work by F.M. has been partially supported by 
IHP-RTN, EC contract No.\ HPRN-CT-2002-00311 (EURIDICE).

\newpage
\section*{Appendix 1: Meson masses from two simulations with larger coupling
($\beta=5.6$) and smaller volume ($16^3\times 48$)}
In this appendix we provide the reader with two sets of results. 
We first give our values equivalent to the ones presented in table~\ref{table2} 
but for $\beta=5.6$. 
Then, for the reader to better appreciate the size of finite volume effects 
at $\beta=5.8$, we also tabulate the pseudoscalar and vector meson masses on the
spatial volumes $16^3$ and $24^3$ in the fully unquenched situations, i.e., 
with $\kappa_v=\kappa_s$.
\begin{table}[h!]
\centering
\begin{tabular}{|c|c|c||c|c|c|}
\hline
$\kappa_{s}$ & $\kappa_{v_1}$ & $\kappa_{v_2}$ & $M_P$ & $M_V$ & 
$m_{v12}^{\rm AWI}(a)$\\
\hline
$0.1560$ & $0.1560$ & $0.1560$ & $0.441(5)$ & $0.541(6)$&$0.0669(4)$\\
  & $0.1560$ & $0.1575$ & $0.398(6)$ & $0.513(8)$&       $0.0548(5)$\\
  & $0.1560$ & $0.1580$ & $0.383(7)$ & $0.503(8)$&       $0.0509(5)$\\
  & $0.1575$ & $0.1575$ & $0.351(7)$ & $0.482(9)$&       $0.0430(5)$\\
  & $0.1575$ & $0.1580$ & $0.334(8)$ & $0.471(10)$&      $0.0392(6)$\\
  & $0.1580$ & $0.1580$ & $0.317(9)$ & $0.459(10)$&      $0.0355(6)$\\
\hline
$0.1575$ & $0.1560$ & $0.1560$ & $0.379(3)$ & $0.456(8)$&$0.0515(4)$\\
  & $0.1560$ & $0.1575$ & $0.333(4)$ & $0.417(13)$&      $0.0394(4)$\\
  & $0.1560$ & $0.1580$ & $0.316(4)$ & $0.403(16)$&      $0.0355(4)$\\
  & $0.1575$ & $0.1575$ & $0.280(4)$ & $0.374(21)$&      $0.0276(3)$\\
  & $0.1575$ & $0.1580$ & $0.261(4)$ & $0.357(27)$&      $0.0238(3)$\\
  & $0.1580$ & $0.1580$ & $0.241(4)$ & $0.340(35)$&      $0.0200(3)$\\
\hline
$0.1580$ & $0.1560$ & $0.1560$ & $0.368(3)$ & $0.468(11)$&$0.0466(4)$\\
  & $0.1560$ & $0.1575$ & $0.324(5)$ & $0.443(15)$&       $0.0348(4)$\\
  & $0.1560$ & $0.1580$ & $0.308(6)$ & $0.436(18)$&       $0.0311(4)$\\
  & $0.1575$ & $0.1575$ & $0.271(7)$ & $0.419(21)$&       $0.0234(5)$\\
  & $0.1575$ & $0.1580$ & $0.250(9)$ & $0.419(27)$&       $0.0199(5)$\\
  & $0.1580$ & $0.1580$ & $0.227(11)$ & $0.424(37)$&      $0.0164(5)$\\
\hline
\end{tabular}
\caption{\small \sl Values of pseudoscalar and vector meson masses, as well as
of the AWI quark masses 
$m_{v12}^{\rm AWI}(a)=\frac{1}{2}\bigl[ m_{v1}(a)+ m_{v2}(a)\bigr]
^{\rm AWI}$ obtained by using eq.~(\ref{AWI}), as obtained from 
the simulation at $\beta=5.6$. All results are given in lattice units.}
\label{tab:m_b56}
\end{table}
\begin{table}[h]
\centering
\begin{tabular}{|c|ccc|ccc|}
\hline
$\kappa_{s}=\kappa_{v}$ & $M_P^{(24)}$ & $M_P^{(16)}$ &$\Delta_{P}$& 
$M_V^{(24)}$ & $M_V^{(16)}$&$\Delta_{V}$ \\
\hline
$0.1535$ & $0.262(4)$ & $0.299(6)$ & $0.14(4)$ & $0.348(8)$ & $0.402(13)$ & 
$0.16(5)$\\
$0.1538$ & $0.236(4)$ & $0.272(10)$ & $0.15(4)$&$0.321(5)$ & $0.356(17)$ & 
$0.11(6)$\\
$0.1540$ & $0.221(3)$ & $0.257(8)$ & $0.17(4)$ & $0.321(9)$ & $0.347(23)$ & 
$0.08(7)$\\
$0.1541$ & $0.182(7)$ & $0.251(9)$ & $0.38(4)$ & $0.291(13)$ & $0.328(31)$ & 
$0.13(8)$\\
\hline
\end{tabular}
\caption{\small \sl Pseudoscalar and vector meson masses, in lattice
units, calculated at $\beta=5.8$ on the lattices $24^3 \times 48$ and
$16^3 \times 48$, respectively. Finite volume effects are quantified by
$\Delta_{P,V} =(M_{P,V}^{(16)}-M_{P,V}^{(24)})/M_{P,V}^{(24)}$.}
\label{tab:m_fv}
\end{table}
\newpage
\section*{Appendix 2: The scale parameter $R_0 = r_0/a$}
\begin{figure}[h]
\begin{center}
\hspace*{-3mm}
\epsfig{file=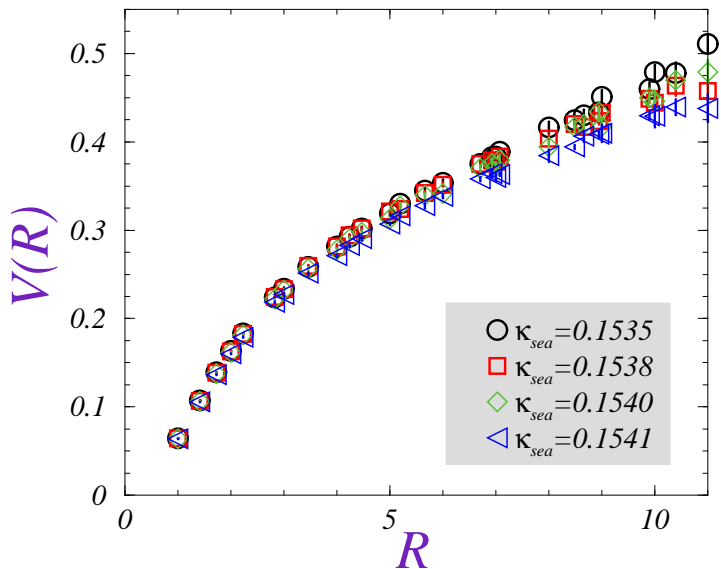, height=8cm}
\vspace*{-0.4cm}
\caption{\small \sl Static quark potential obtained from the simulation with 
$N_f=2$ at $\beta=5.8$ on the lattice $24^3\times 48$.\label{fig:V}  } 
\end{center}
\end{figure}

To determine the Sommer scale parameter $R_0$~\cite{sommer} we computed the 
static quark potential $V(R)$ which is extracted from the time
dependence of the rectangular Wilson loops $W(R,t)$, where $R$ is the spatial 
separation between the static charges, i.e., 
\be
W(R,t)= {\cal C}(R) e^{-V(R)t}\,\Rightarrow\,V(R)=\lim_{t\to \infty}V_{\rm
eff}(R,t)\equiv\lim_{t\to \infty} \log\left({W(R,t)\over W(R,t+1)}\right)\,.
\ee
Large time separations are reached after using the so-called HYP (hypercubic 
blocking) procedure proposed in ref.~\cite{knechtli} which helps in improving 
the signal for the Wilson loops. 
In this way we find a stable signal for $V(R)$ for $t_{\rm fit} \in [8,13]$. 
At intermediate spatial separations between the static charges, we then fit the 
data (shown in fig.~\ref{fig:V}) to the form
\be
\label{eq1}
V(R) = V_0 + \sigma R- {e\over R}  - g \delta V(R)\,,
\ee
where we also account for $\delta V(R)$, the perturbatively estimated 
discretisation correction to the Coulomb potential~\cite{rebbi}. 
The Sommer scale $R_0$ is defined as
\be
\biggl(R^2 {d V(R)\over d R}\biggr)_{R=R_0} = 1.65\,.
\ee
The fitting window consistent with the form~(\ref{eq1}) is found for 
$R\in [2,7]$ after which the statistical quality of the data for $V(R)$ rapidly 
deteriorates. 
From the fit of our data to the form~(\ref{eq1}) in that window, we obtain 
\bea\label{eq0}
\beta=5.8\,:\qquad\kappa_s&=&\{0.1535,\,0.1538,\,0.1540,\,0.1541\}\,,\nn\\
R_0 &=& \{7.52(17),\,7.70(9),\,7.78(19),\,8.10(20)\}\,,\nn\\
\sqrt{\sigma} &=& \{0.155(4),\,0.151(2),\,0.150(4),\,0.144(4)\}\,,\\
e &=& \{0.289(10),\,0.302(5),\,0.294(7),\,0.288(13)\}\,,\nn\\
V_0&=&\{0.260(7),\,0.268(3),\,0.264(5),\,0.262(9)\}\,,\nn\\
g&=&\{0.033(14),\,0.036(9),\,0.035(7),\,0.015(15)\}\,,\nn
\eea
ordered as the values of $\kappa_s$. Our results for $R_0$ obtained at the same 
value of $\beta$ but on the smaller volume ($16^3\times 48$) agree with those 
written in eq.~(\ref{eq0}), within the statistical errors. 

At $\beta=5.6$, we find stability for $t_{\rm fit}\in[3,5]$, and then from 
the fit to  eq.~(\ref{eq1}) in $R\in [2,7]$, we have
\bea
\beta=5.6\,:\qquad\kappa_s&=&\{0.1560,\,0.1575,\,0.1580\}\,,\nn\\
R_0 &=& \{5.15(5),\,5.72(10),\,5.98(8)\}\,,\nn\\
\sqrt{\sigma} &=& \{0.225(3),\,0.202(4),\,0.195(3)\}\,,\\
e &=& \{0.305(11),\,0.317(8),\,0.308(6)\}\,,\nn\\
V_0&=&\{0.262(9),\,0.279(7),\,0.275(5)\}\,,\nn\\
g&=&\{0.009(11),\,0.022(7),\,0.020(4)\}\,,\nn
\eea
in good agreement with ref.~\cite{orth}.

\end{document}